\documentstyle[12pt]{article}
\newcommand \bq {\begin{eqnarray}}
\newcommand \be {\begin{equation}}
\newcommand \eq {\end{eqnarray}}
\newcommand \ee {\end{equation}}

\newcommand \ran {\rangle}
\newcommand \0 {|0 \rangle}
\newcommand \io {\langle 0|}
\newcommand \1 {|1 \rangle}
\newcommand \kb {| \bar 0 \rangle}
\newcommand \ikb {\langle \bar 0|}
\newcommand \zo { \bar 0}
\newcommand \lan {\langle}

\newcommand \si {\sigma }

\newcommand \th {\theta}
\newcommand \r {\rho}
\newcommand \rtil { \tilde \rho}

\begin{document}

\setcounter{page}{1}
\vspace{2cm}
\begin{center}
\vspace{5mm}
\vspace{5mm} {\large The broadcast quantum channel for classical information transmission.
}\\
\vspace{5mm} {\large  A.E. Allahverdyan, D.B. Saakian }\\
\vspace{5mm}
{\em Yerevan Physics Institute}\\
{\em Alikhanian Brothers St.2, Yerevan 375036, Armenia}\\
\end{center}

\vspace{5mm}
\centerline{\bf{Abstract}}

It is well known that quantum theory forbids the exact copying of an unknown quantum state. 
Therefore 
in broadcasting of classical information by a quantum channel  
an additional contribution to the error in the decoding is expected. We consider the 
optimal copying transformation  which is adapted to classical information transmission 
by two linearly independent quantum states, and show that there is no additional contribution to 
the error. Instead the clones are correlated, and this breaks their usefulness: The entanglement
increases with non-orthogonality of the states.
The capacity of the corresponding quantum channel  is considered also.

\newpage
\vspace{8mm}

Exact quantum copying (cloning) of an unknown quantum state does not exist. 
This well known theorem of Wootters and Zurek
\cite{zurek} has been recently generalized to mixed states \cite{barnum}. 
The physical origin of this result is the following:
states from an arbitrary set cannot be distinguished from each other, 
therefore the complete information cannot be obtained.
Thus {\it quantum information} cannot be cloned perfectly. In this paper we wish to answer to the
related question : is it possible to clone {\it quantum} states which represent {\it classical information}?

The problem is the following. The classical information is represented as  a sequence of bits 
(i.e. $0$, $1$ symbols).
We propose that in a  long sequence the frequencies of $0$ and $1$ are the same, 
i.e. the {\it a priori} probabilities are
equal.  A {\it quantum coder} generates the states $\0 $ ($\1 $) for the symbol $0$ ($1$)
\be 
\label{1}
\lan 1\0 =\cos \th ,\ \ 0\leq \th \leq \frac{\pi }{2}
\ee
This nonorthogonality can be connected with the construction of the coder 
(which generates the coherent states for example). 
On the other hand nonorthogonal states can be more optimal
if a quantum channel is subjected to external noise; also they are used in quantum cryptography
\cite{crypto}. In the second step the quantum states should be cloned and 
broadcasted to the users A, B. Thus the whole
scheme is the following
\begin{equation}
\label{2}
i\mapsto |i\ran \mapsto \hat U(|{\rm blank}\ran |i\ran)=\si _i\mapsto
\begin{array}{cc} \r _i={\rm tr}_s \si _i \mapsto {\rm user \ \ A}
               \\ \rtil _i={\rm tr}_b \si _i \mapsto {\rm user \ \ B} 
\end{array}
\end{equation}
Where $i=0,1$, $\hat U$ is a cloning transformation, 
$|{\rm blank}\ran $ is a "blank" state of the machine, 
${\rm tr}_s $, $ {\rm tr}_b $ are  partial traces by the subspaces of the system and the "blank". 
Because {\it classical} information is 
transmitted we should optimize the error of {\it distinguishability} between $\r _1$ and $\r _0$, 
as well as between $\rtil _1$ and $\rtil _0$. Further, we shall choose $\r _i=\rtil _i$ and $|{\rm blank}\ran =\0$.
Also we have considered only unitary $\hat U$ in the corresponding
four dimensional Hilbert space. Because the users are 
equivalent we should minimize  the average error in the distinguishing between $\r _0$ and $\r _1$. If the user makes POVM
measurement $\Pi _1+\Pi _0=1$, then the average error is the following (recall that a priori probabilities are the same)
\be
\label{3}
P_e=\frac{1}{2}p(0/1)+ \frac{1}{2}p(1/0)=\frac{1}{2}(1+{\rm tr}(\r \Pi _1)),\  \ \r =\r _0-\r _1
\ee
We see that ${\rm tr}(\r \Pi _1)=\sum_i(\r \Pi _1)_{(i)}$ should be minimized (where $A_{(i)}$ is a corresponding eigenvalue of
the matrix $A$). 
Thus the resulting formula is 
\cite{helstrom}
\bq
\label{5}
& & \Pi _1 ^{({\rm opt})}=\sum_i\th (-\r _{(i)})|\r _{(i)}\ran \lan \r _{(i)}| \nonumber \\
& & \Pi _0 ^{({\rm opt})}=\sum_i\th (\r _{(i)})|\r _{(i)}\ran \lan \r _{(i)}| \nonumber \\
& & P _e ^{({\rm opt})}=\frac{1}{2}(1+\sum_i\th (-\r _{(i)})\r _{(i)})
\eq
For density matrices in a two dimensional Hilbert space the simpler formula can be obtained
\be
\label{6}
P _e ^{({\rm opt})}=\frac{1}{2}(1+\r _{({\rm min})})
\ee
Where $\r _{({\rm min})}$ is the minimal eigenvalue of $\r $. So our cloning machine should minimize (\ref{6}).\\
The unitary operation $\hat U$ can be written as
\bq
\label{7}
&& \hat  U(\0 _s\0 _b)=a_0\0 _s\0 _b+b_0\0 _s\kb _b+c_0\kb _s\0 _b+d_0\kb _s\kb _b \nonumber \\
&& \hat  U(\kb _s\0 _b)=a_{\zo }\0 _s\0 _b+b_{\zo }\0 _s\kb _b+c_{\zo }\kb _s\0 _b+d_{\zo }\kb _s\kb _b
\eq
Where $\lan 0\kb =\delta _{0\zo }$. We can also write
\be
\label{8}
\hat  U(\1 _s\0 _b)=a_1\0 _s\0 _b+b_1\0 _s\kb _b+c_1\kb _s\0 _b+d_1\kb _s\kb _b 
\ee
For the conditions $\rtil _0=\r _0$ and $\rtil _1=\r _1$ is sufficient to choose 
$b_0=c_0$ and $b_1=c_1$. 
Further we propose that all coefficients in (\ref{7}, \ref{8}) are real. Now (\ref{6}) can be written 
as
\be
\label{9}
P_e=\frac{1}{2}(1-\sqrt{\Lambda }),\ \ \Lambda =
(a_1c_1+b_1d_1-a_0c_0-b_0d_0)^2+a_1^2+b_1^2-a_0^2-b_0^2
\ee
So $\Lambda $ should be maximized with the constraints
\be
\label{10}
a_0^2+2b_0^2+d_0^2=1,
\ee
\be
\label{11}
a_1^2+2b_1^2+d_1^2=1,
\ee
\be
\label{12}
a_1a_0+2b_1b_0+d_1d_0=\cos \th .
\ee
The problem of maximization can be simplified by introducing the new variables
\be
\label{13}
a_{1(0)}=\frac{1}{\sqrt{2}}(x_{1(0)}+y_{1(0)}),\  \
d_{1(0)}=\frac{1}{\sqrt{2}}(x_{1(0)}-y_{1(0)})
\ee
After some calculations (which are not reproduced here) we get the final result:
for the optimal unitary transformations like
\bq
\label{17}
&& a_{0(1)}=\frac{1}{\sqrt{2}}(\pm \sin \frac{\th }{2 }+\cos \frac{\th }{2 }\cos \phi )
\nonumber \\
&& d_{1(0)}=-a_{0(1)} \nonumber \\
&& b_1=b_0=\frac{1}{\sqrt{2}} \cos \frac{\th }{2 }\sin \phi
\eq
(here $0\leq \phi \leq 2\pi $ is the free parameter) {\it the error of distinguishing
between $\r _1$ and $\r _0$ is the same as for $\1 $ and $\0 $}:
\be
\label{177}
P_e=\frac{1}{2}(1-\sin \th )
\ee
It is of course the maximal value which is possible in a cloning  because 
after action of any quantum transformation (it can be represented as an unitary transformation 
plus partial trace) quantum states cannot be more distinguishable (this  fact can be checked by 
simple calculation ).
For the marginal density matrices we have
\be
\label{18}
 \r _{0(1)}= \frac{1}{2}(1\pm \sin \th \cos \phi )\0 \io +\frac{1}{2}(1\mp \sin \th \cos \phi )\kb \ikb 
\pm \frac{1}{2}\sin \th \sin \phi (\0 \ikb + \kb \io) 
\ee
The optimal measurement can be written as
\bq
\label{d19}
&& \Pi _{0(1)}=|\psi _{\pm }\rangle \langle \psi _{\pm }| \nonumber \\
&& |\psi _{\mp }\rangle =\frac{1}{\sqrt{2(1\mp \cos \phi)}}(
(\mp 1 +\cos \phi )\0 +\sin \phi \kb )
\eq
Two pure states ina Hilbert space of any dimension span only a two-dimensional subspace;
hence {\it any two nonorthogonal pure state can be cloned without additional error in the decoding}.

By (\ref{1}) we see that the copies are in entanglement state. 
Really this problem can be very important
because by working with the first copy the second is changed also. 
Because the clones are the subsystems of the pure system 
as a measure of entanglement can be used the quantum entropy of a marginal density matrix  
\cite{ben} (there are many possible measures, we choose that
which is more convenient for us):
\be
\label{20}
S=-{\rm tr}\sigma \ln \sigma
\ee
For pure state (\ref{20}) is zero, and it is positive function for all other cases. 
As it should be, $S$ is maximized
with uniform distribution. Now for the entanglement of our clones we have
\bq
\label{21}
&&S(\r _0)=S(\r _1)=h(P_e), \nonumber \\
&& h(x)=-x\ln x -(1-x)\ln (1-x)
\eq
We see  that {\it entanglement is maximal in the "worst" case
when}  $ \th \mapsto \pi /2$.

In the last part of the paper we discuss the qualitative measures of information 
transmission through  broadcast channels.
We discuss only one possible, practically important scenario \cite{coverelgammal}: 
when two independent classical 
sources communicate by the same 
generator of quantum states (coding machine) with the users A and B (correspondingly)
{\it at the same time}. We start with the general theory \cite{coverelgammal}, 
and after this apply it 
to our case.

Suppose  that possible quantum states
of the coding machine have a priori probabilities $p_x$.
After action of a cloning transformation and a measurement the user A (B)
obtain a classical message $y$ ($z$) with a  probability $p_y$ ($p_z$)
(the role of possible entanglement here will be discussed later):
\be
\label{gg3}
p_z=\sum_xp_1(z/x)p_x, \  \ p_y=\sum_xp_2(y/x)p_x,
\ee
where a noise is described by the sets of conditional probabilities 
\be
\label{dop1}
p_1(y/x), \  \ p_2(z/x)
\ee
Let us denote the  ensemble of states for coding machine by $X$, and the messages of the user
A(B) are in the ensemble $Y$($Z$). 

Now for $N_1$($N_2$) symbols of 
the source $1(2)$  $N$ states of the ensemble $X$ is generated and 
transmitted to the users. 
The users must separate and recognize their messages because the 
user A(B)
wants to have only messages from the source $1$($2$).
\footnote{The applications of this scheme to real life (for example in TV) are
discussed in \cite{coverelgammal}} 
In this sense each source acts as a noise for the other, 
so $R_1=N_1/N$ and $R_2=N_2/N$ are supplemented. Now there are three different sources 
for the noise in this channel: initial non-orthogonality  of the coding machines states, 
non-orthogonality which can occur after action of a cloning transformation,
and the noise which is introduced by one user  for other.
Therefore
must be $N>N_1, N_2$ because "redundancy against a noise" should be ensured. 

Thus the broadcast channel is defined by the ensembles $X$, $Y$, $Z$, the a priori probabilities
$p_x$, and the conditional probabilities $p_1(y/x)$, $p_2(z/x)$.

Now we assume that the channel is 
{\it degraded}:
\be
\label{27}
p_2(z/x)=\sum_yW(z/y)p_1(y/x), \ \ W(z/y)\geq 0, \ \ \sum_zW(z/y)=1
\ee
This means that the transmission scheme can be formally represented as
\be 
\label{28}
X\underbrace{ \longmapsto }_{{\rm 1- channel}}Y\underbrace{ \longmapsto }_
{{\rm W- channel}}Z
\ee
Or even in the more general form
\be
\label{29}
S\underbrace{ \longmapsto }_{{\rm 0- channel}}X\underbrace{ \longmapsto }_{{\rm 1-channel}}Y
\underbrace{ \longmapsto }_{{\rm W-channel}}Z
\ee
Where W-channel is described by $W(z/y)$, and 0-channel is introduced as "{\it trade-off channel}" 
between $R_1$ and $R_2$.
The following result has been obtained in \cite{coverelgammal}:
{\it for reliable connection (i.e. a connection with small probability of the error in the decoding) between 
the sources and their addresses should be}
\be
\label{30}
R_1\leq I(X:Y/S), \ \ R_2\leq I(S:Z)
\ee
Where (we measure the information functions in nuts)
\begin{equation}
\label{24d}
I(X:Y /S )=\sum_{x ,y ,s }
p(x ,y ,s)\ln \frac{p(y /x s
)}{p(y /s)},
\end{equation}
\begin{equation}
\label{25d}
I(S :Z )=\sum_{s ,z}
p(s,z )\ln \frac{p(s /z
)}{p(s)}.
\end{equation}
The second value is usual mutual information between the ensembles 
$S$ and $Z$.
The first  value is called mutual-conditional information
(mc-information). The mutual information of
two ensembles is the reduction of entropy of
one ensemble if the other is observed.
Mc-information has the same meaning but after realization of
the conditional ensemble (i.e. $S$ in our case).
The physical meaning of (\ref{24d}, \ref{25d}) can be understood from the eq. (\ref{29}):
$R_2$ is determined by the direct connection between $S$ and $Z$, for determination of 
$R_1$ the ensemble $S$ should be fixed. If 0-channel is out (totally noised)
then $R_1$ ($R_2$) is maximal (minimal),
and if 0-channel is noise-free then the opposite case is realized: $R_2$ ($R_1$) is maximal (minimal).

Now we apply this theory to our problem:
The states of coding machine are $|0\rangle$ and $|1\rangle$ with the equal a priori probabilities, as
a cloning transformation we use (\ref{17}), and the users 
for obtaining their classical messages make the same measurement (\ref{d19}). The resulting channel
is degraded, and $W(z/y)=\delta _{zy}$.

As we have seen the entanglement is introduced by 
the cloning transformation. Thus the users are dependent, and the distributions (\ref{dop1}) are 
marginally distributions of the more general distribution $p(yz/x)$. 
Fortunately, the capacities of a degraded
broadcast channel depend only from the marginal distributions (\ref{dop1}) \cite{coverelgammal}.

We assume that 0-channel is memory-less, has equal a priori probabilities (it can be shown that this 
choice is optimal), 
and the following formula
for the '0-noise' is holds
\be
\label{dop2} 
p_0(0/1)=p_0(1/0)=\epsilon
\ee
The final results are the following
\bq
\label{dopinf}
I(X:Y/S)&=&h((1-P_e)\epsilon +P_e(1-\epsilon ))-h(P_e) \nonumber \\
I(S:Z)&=&\ln 2-h((1-P_e)\epsilon +P_e(1-\epsilon ))
\eq
It is sufficient to assume that $0\leq \epsilon \leq 0.5$, if $\epsilon =0(0.5)$
then $I(S:Z)$ $(I(X:Y/S))$ is maximal, and the the opposite quantity is minimal.
It is remarkable that the capacities of the broadcast quantum channel depend only from $P_e$-
the error of the decoding between the two initial quantum states.


We consider the cloning machine which is adapted to classical information transmission,
and show that the cloning introduces the entanglement but there is no an additional 
contribution to the error in the decoding. The entanglement increase with indistinguishability 
of the initial quantum states.

We also computed
the capacities of the corresponding broadcast channels channels. 
There is further work to be done. We think
that the most important problem in this direction is to consider the broadcast 
channels when the information is
transmitted by quantum states of an electromagnetic field.


\begin{thebibliography}{99}

\bibitem{zurek} W.K. Wootters and W.H. Zurek, Nature, {\bf 299}, 802, (1982).

\bibitem{barnum} H.Barnum et al, Phys.Rev.Lett., {\bf 76}, 2818, (1996).





\bibitem{crypto} R.J. Hughes, et all., Contemporary Physics, {\bf 36}, 149, (1995).

\bibitem{helstrom}C.W. Helstrom, {\it Quantum detection
and estimation  theory.} Academic Press, 1976.

\bibitem{strat}R. L. Stratonovich,
{\it Information Theory}, Moscow, Nauka 1975.

\bibitem{coverelgammal}A. El Gammal, T. Cover, Proc. IEEE, {\bf
68},1466,(1980).

\bibitem{ben} C.~H. Bennett, D.~P. DiVincenzo, J.~A. Smolin,
W.~K. Wootters,  Phys. Rev. A {\bf54}, 3824 (1996); e-print
quant-ph/9604024.






\end{thebibliography}
\end{document}